%% file: talk.tex
\documentstyle[times,pramana,floats]{ias}
\begin{document}
\input epsfa.tex

\epsfclipon
\def\sfrac#1#2{{\textstyle\frac{#1}{#2}}}
\newcommand{\lsim}{\mbox{\raisebox{-.6ex}{~$\stackrel{<}{\sim}$~}}}
\newcommand{\gsim}{\mbox{\raisebox{-.6ex}{~$\stackrel{>}{\sim}$~}}}
\mark{{Mimicking transPlanckian effects in the CMB with conventional physics}{J.M.Cline}}
\title{Mimicking transPlanckian effects in the CMB\\ with conventional physics}

\author{J.M.Cline}
\address{Theory Division, CERN, CH-1211, Geneva 23, Switzerland and\\
Physics Department, McGill University,
Montr\'eal, Qu\'ebec, Canada H3A 2T8}
\abstract{ We investigate the possibility that fields coupled to the
    inflaton can influence the primordial spectrum of density
    perturbations through their coherent motion. For example, the
    second field in hybrid inflation might be oscillating at the
    beginning of inflation rather than at the minimum of its potential.
    Although this effect is washed out if inflation lasts long enough,
    we note that there can be up to 30 e-foldings of inflation prior to
    horizon crossing of COBE fluctuations while still giving a
    potentially visible distortion. Such pumping of the inflaton
    fluctuations by purely conventional physics can resemble
    transPlanckian effects which have been widely discussed. The
    distortions which they make to the CMB could leave a distinctive
    signature which differs from generic effects like tilting of the
    spectrum.
}

\maketitle
\section{Motivation}
There has been much discussion of the possible effects of transPlanckian
physics on inflationary perturbations recently.  Typically density
perturbations which have macroscopic wavelengths today could have been
shorter than the Planck distance at sufficiently early times, when
inflation was taking place.  It is therefore conceivable that new physics
could leave its imprint on the CMB.  In this work \cite{us} we investigate an
alternative scenario, that some field coupled to the inflaton could
distort the primordial inflaton fluctuations due to its coherent motion.
The effects could be qualitatively similar to those of the more exotic
transPlanckian possibilities.  

\section{TransPlanckian physics} 
It has been argued that transPlanckian physics would manifest itself by putting
the inflaton in a vacuum state which differs from the usual Bunch-Davies vacuum.
We recall that the latter is defined to be the one annhilated by the operator 
coefficient of the positive frequency solution $\phi^+_k$,
$\phi(t,x) = \sum_k\left( a_k \phi^+_k  + a_k^\dagger \phi^-_k \right)$
where $\phi^\pm_k = N_k\left(\pm iH + {k\over a}\right) e^{\pm ik/aH + ikx}$ for
a massless inflaton.  The alternative states which have been considered, $\alpha$-vacua,
are related to the standard vacua by a Bogoliubov transformation, $\phi_k^+ = 
A_\alpha \tilde\phi_k^+ + B^*_\alpha\tilde\phi_k^- $.  The effect on the power spectrum
of density perturbations is a modulation, $\tilde P(k) \sim 
|\tilde\phi_k^+/\phi_k^+|^2 P(k)$, where the wave functions are to be evaluated at
$t=\infty$.  It has been argued \cite{ulf} that the modulation factor should have the form
$1-{H\over\Lambda}\sin({2\Lambda\over H})$, where $H$ is evaluated at the time
when the mode $k$ crosses the horizon.  If $H$ varies fast enough, the effect would
be an oscillatory modulation as a function of $k$ \cite{shiu}.

\section{Conventional physics} An oscillatory modulation of the power spectrum can also
be produced if the inflaton $\phi$ couples to an oscillating field $\chi$.  This could
naturally happen in hybrid inflation models, with potential $V(\phi,\chi) = \frac12 \,
m^2 \, \phi^2 + \lambda (\chi^2 - v^2)^2 + \frac12 \, g \, \chi^2 \phi^2 +  \tilde\lambda
\, \phi^4$, if $\chi$ starts far away from its ground state.  Ordinarily one neglects
such a possibility since sufficient inflation will quickly damp out the oscillations. But
if there is only a limited amount of inflation prior to horizon-crossing of COBE-scale
modes, then $\chi$ could still be oscillating when CMB fluctuations are being produced.
The inflaton fluctuation equation of motion is 
\begin{equation}
\label{phieq}
\ddot\varphi_k + 3H\dot\varphi_k + k^2e^{-2Ht} \varphi_k = {g \chi^2(t)\varphi_k}
\end{equation} 
Since the amplitude of the oscillations
redshifts like $a(t)^{-3/2}$, we can estimate how many how many $e$-foldings of inflation
are needed to render $\chi$ ineffective.  By comparing the new term in the equation of
motion with $k^2e^{-2Ht}$ at the time of horizon crossing (when $k/a = H$, we find the
maximum number of $e$-foldings before horizon-crossing to be $ N_e \ =\
\ln\left({a/a_0}\right) \  =\ \sfrac13\ln\left( {g\chi_0^2/ H^2}\right)$.  Demanding that
$\chi$ not dominate the energy density (which would give matter domination instead of
inflation), we find the constraint $N_e \ <\  \sfrac13\ln\left({3gM_p^2/M^2}\right) \
\lsim\ 27$ (taking $g=1$, $M=100$ GeV), where $M$ is the effective mass of $\chi$ during
inflation.  A more careful estimate which also demands that
$\chi$ is not damped too quickly by producing inflatons gives $N_e\lsim 12$.  We will
also consider an alternative model in which the inflaton-$\chi$ coupling is 
$g'\chi\phi^2$ instead of $g\chi^2\phi^2$.  This gives 
$N_e\lsim 30$.

\begin{figure}[b]
\centerline{\epsfxsize=0.5\textwidth\epsfbox{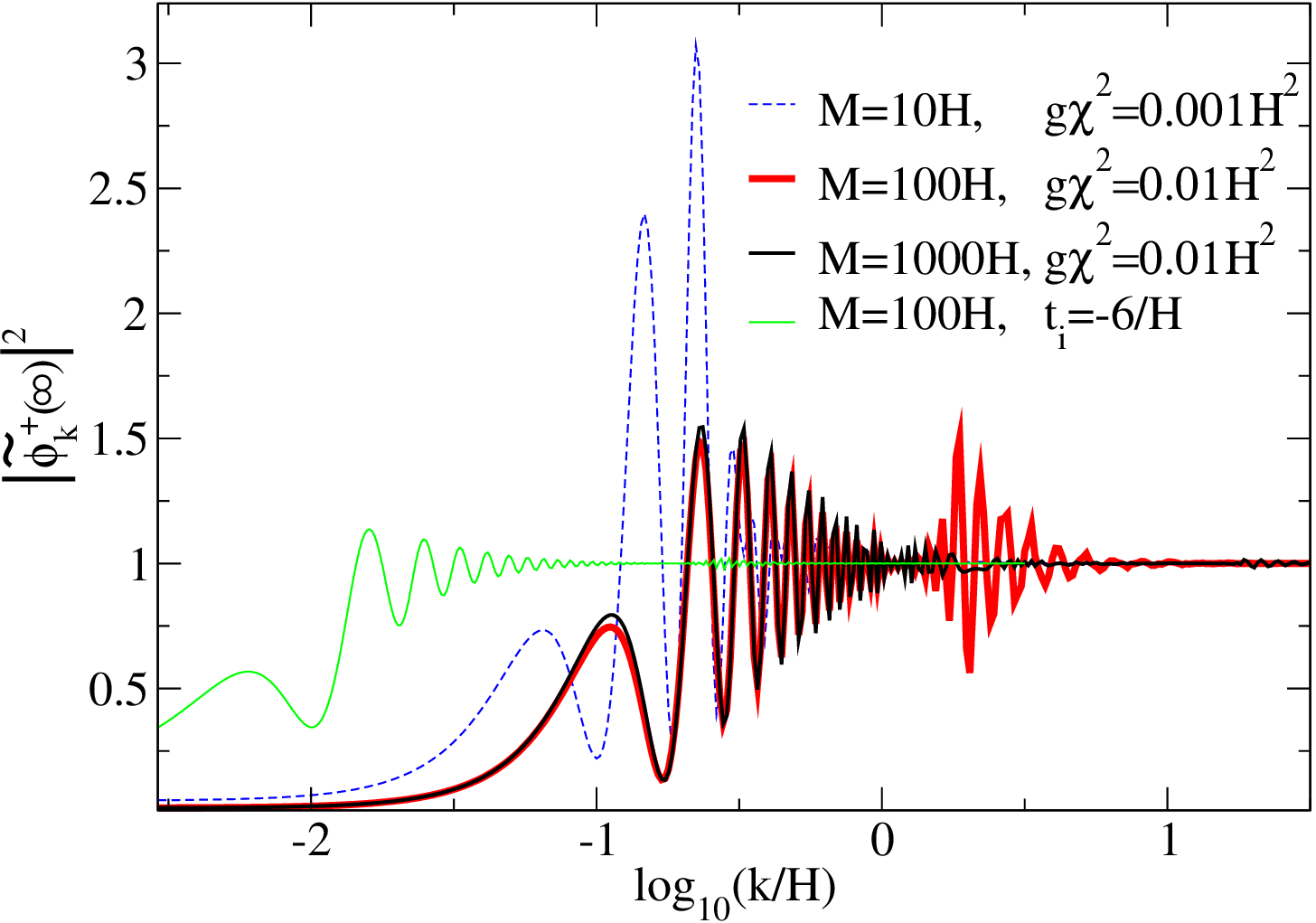}\epsfxsize=0.5\textwidth\epsfbox{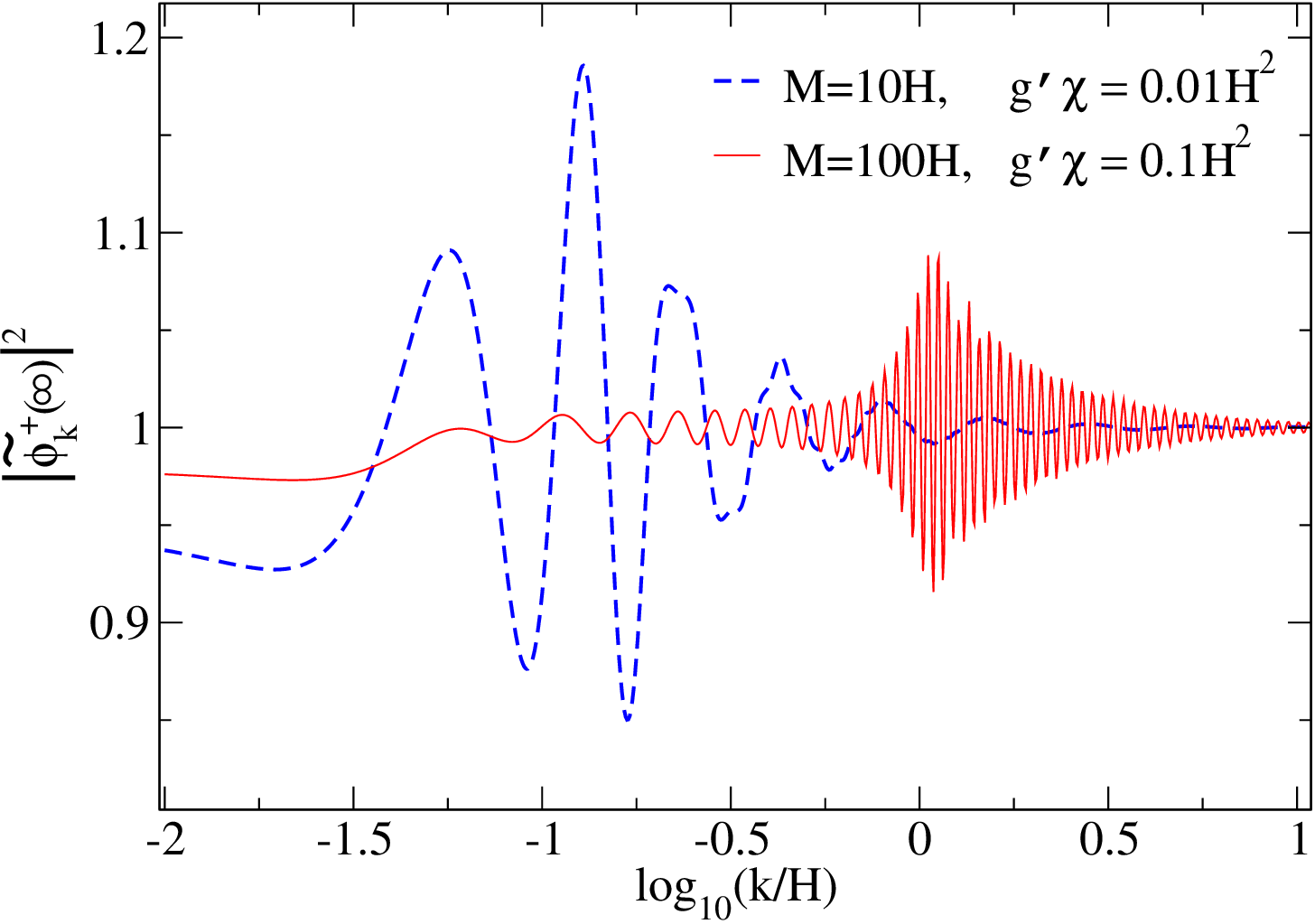}}
\caption{$P(k)$ versus $\log(k)$ in the $g\chi^2\phi^2$ model (left) and the $g'\chi\phi^2$ model (right),
for several values of the $\chi$ mass $M$ and amplitude at horizon crossing.}
\label{fig:x}
\end{figure}

To find the effect of $\chi$ oscillations on the inflaton fluctuations, we solve
(\ref{phieq}) both numerically and partially analytically using the Greens' function
method; both give useful insights.  Figure 1 shows the deviations in $P(k)$ for examples
of both the $g\chi^2\phi^2$ and $g'\chi\phi^2$  models, from numerical solutions.  An
interesting difference between the two models is that power is strongly suppressed at
low $k$ in the first, whereas the maximum deviation occurs at intermediate $k$ values 
in the second.   From the analytic approach can compute the maximum size of the 
fractional deviation in $P(k)$:
\begin{equation}
{\delta P\over P}_{\rm max} = 
{g \chi_{0}^2 \over 9 H^2}\ (g\chi^2\phi^2\hbox{\ model}),\quad {g' \chi_{0}
   \over H^{1/2} (M/2)^{3/2}}\ (g'\chi\phi^2 \hbox{\ model})
\end{equation}
Interestingly, these are not suppressed by powers of $H$ as is the case with
transPlanckian or effective field theory effects \cite{kaloper}; rather they are enhanced by small
$H$.

\section{Effect on CMB}
To make contact with observations, we have input our primordial spectra into the 
CMBFAST code \cite{CMBFAST} to generate the resulting temperature anisotropy (Doppler peaks).  Some results are shown in
fig.\ 2.  The $g\chi^2\phi^2$ model tends to give the largest deviations at low
multipole values, whereas the $g'\chi\phi^2$ model allows for maximal deviations
at higher $l$ values.  The latter possibility is interesting for current and future
experiments which will be sensitive to smaller angular scales, hence larger $l$.

\begin{figure}[h]
\centerline{\epsfxsize=0.5\textwidth\epsfbox{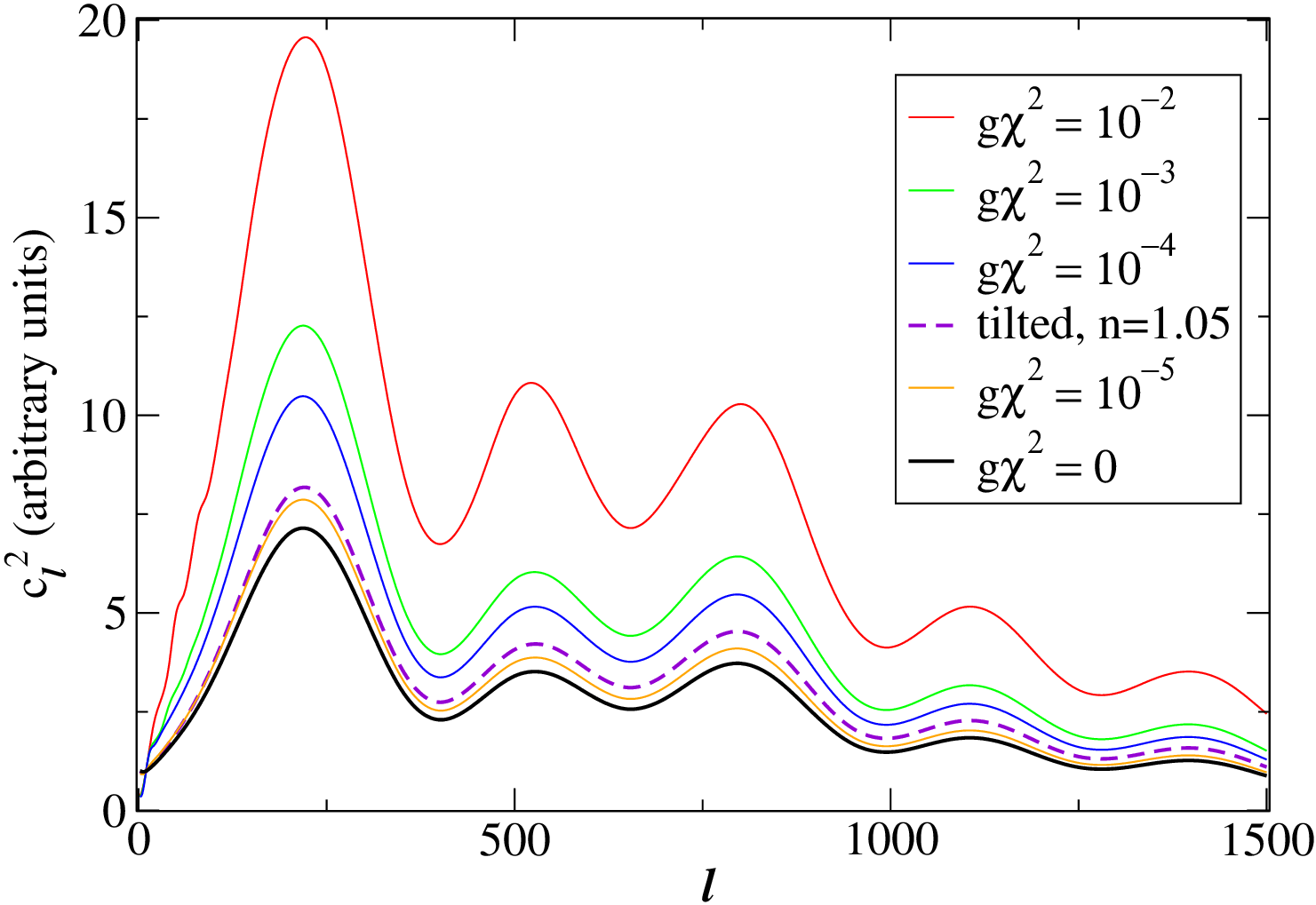}
\epsfxsize=0.5\textwidth\epsfbox{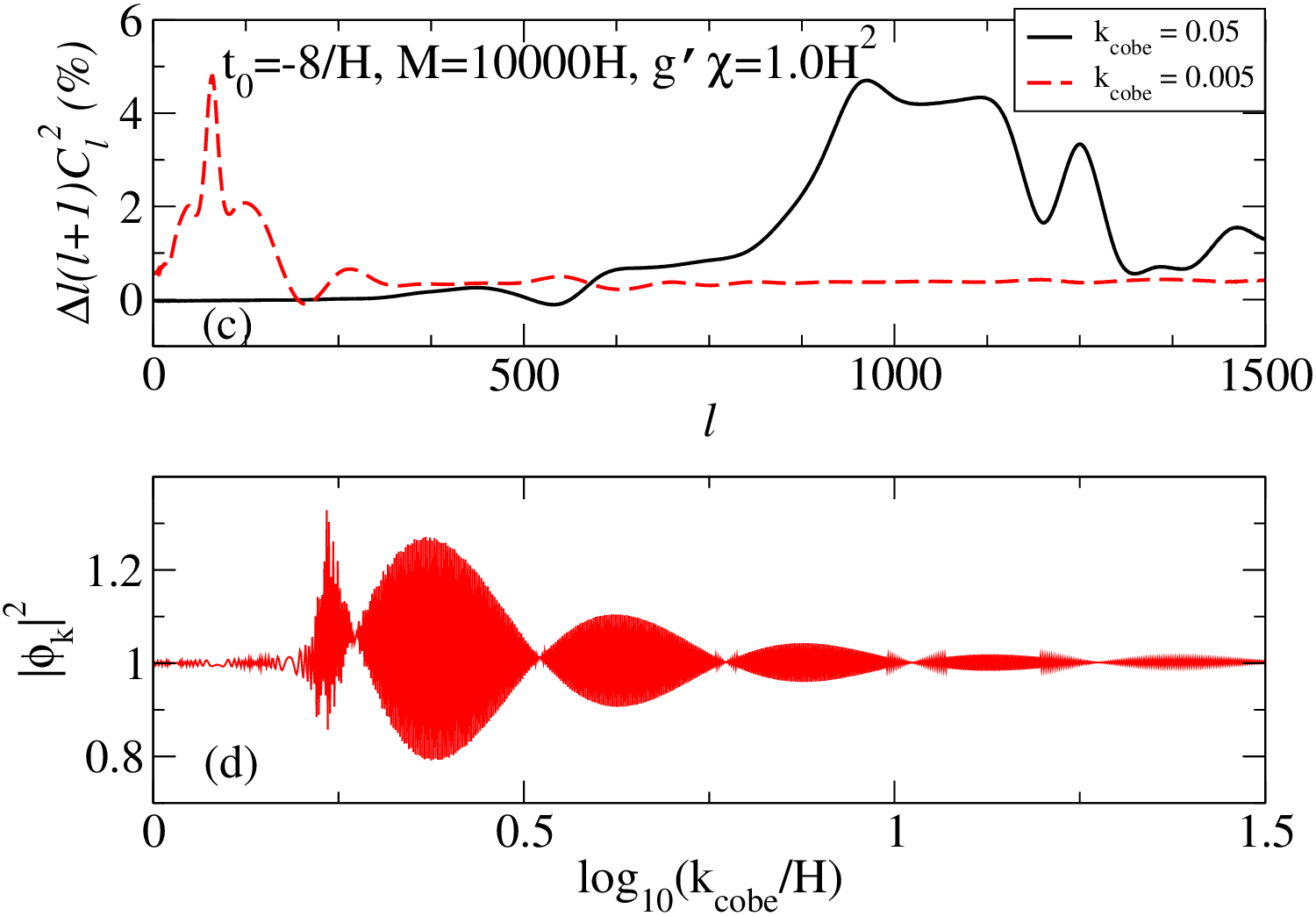}}
\caption{Left: distortion of temperature anisotropy in $g\chi^2\phi^2$ model, for different
values of $g\chi^2$.  Right: percent deviation in temperature anistropy in $g'\chi\phi^2$ model
(upper panel) and the underlying inflaton power spectrum (lower panel).}
\label{fig:y}
\end{figure}

\end{document}

%% file: epsfa.tex
\ifx\epsfannounce\undefined \def\epsfannounce{\immediate\write16}\fi
 \epsfannounce{This is `epsf.tex' v2.7k <10 July 1997>}%
\newread\epsffilein    
\newif\ifepsfatend     
\newif\ifepsfbbfound   
\newif\ifepsfdraft     
\newif\ifepsffileok    
\newif\ifepsfframe     
\newif\ifepsfshow      
\epsfshowtrue          
\newif\ifepsfshowfilename 
\newif\ifepsfverbose   
\newdimen\epsfframemargin 
\newdimen\epsfframethickness 
\newdimen\epsfrsize    
\newdimen\epsftmp      
\newdimen\epsftsize    
\newdimen\epsfxsize    
\newdimen\epsfysize    
\newdimen\pspoints     
\pspoints = 1bp        
\epsfxsize = 0pt       
\epsfysize = 0pt       
\epsfframemargin = 0pt 
\epsfframethickness = 0.4pt 
\def\epsfbox#1{\global\def\epsfllx{72}\global\def\epsflly{72}%
   \global\def\epsfurx{540}\global\def\epsfury{720}%
   \def\lbracket{[}\def\testit{#1}\ifx\testit\lbracket
   \let\next=\epsfgetlitbb\else\let\next=\epsfnormal\fi\next{#1}}%
%
%
\def\epsfgetlitbb#1#2 #3 #4 #5]#6{%
   \epsfgrab #2 #3 #4 #5 .\\%
   \epsfsetsize
   \epsfstatus{#6}%
   \epsfsetgraph{#6}%
}%
\def\epsfnormal#1{%
    \epsfgetbb{#1}%
    \epsfsetgraph{#1}%
}%
\newhelp\epsfnoopenhelp{The PostScript image file must be findable by
TeX, i.e., somewhere in the TEXINPUTS (or equivalent) path.}%
\def\epsfgetbb#1{%
%
%
    \openin\epsffilein=#1
    \ifeof\epsffilein
        \errhelp = \epsfnoopenhelp
        \errmessage{Could not open file #1, ignoring it}%
    \else                       
        {
            \chardef\other=12
            \def\do##1{\catcode`##1=\other}%
            \dospecials
            \catcode`\ =10
            \epsffileoktrue         
            \epsfatendfalse     
            \loop               
                \read\epsffilein to \epsffileline
                \ifeof\epsffilein 
                \epsffileokfalse 
            \else                
                \expandafter\epsfaux\epsffileline:. \\%
            \fi
            \ifepsffileok
            \repeat
            \ifepsfbbfound
            \else
                \ifepsfverbose
                    \immediate\write16{No BoundingBox comment found in %
                                    file #1; using defaults}%
                \fi
            \fi
        }
        \closein\epsffilein
    \fi                         
    \epsfsetsize                
    \epsfstatus{#1}%
}%
%
\def\epsfclipon{\def\epsfclipstring{ clip}}%
\def\epsfclipoff{\def\epsfclipstring{\ifepsfdraft\space clip\fi}}%
\epsfclipoff 
%
%
\def\epsfspecial#1{%
     \epsftmp=10\epsfxsize
     \divide\epsftmp\pspoints
     \ifnum\epsfrsize=0\relax
       \includegraphics{\ifepsfdraft}%
     \else
       \epsfrsize=10\epsfysize
       \divide\epsfrsize\pspoints
       \includegraphics{\ifepsfdraft}%
     \fi
}%
%
\def\epsfframe#1%
{%
  \leavevmode                   
  \setbox0 = \hbox{#1}%
  \dimen0 = \wd0                                
  \advance \dimen0 by 2\epsfframemargin         
  \advance \dimen0 by 2\epsfframethickness      
  \vbox
  {%
    \hrule height \epsfframethickness depth 0pt
    \hbox to \dimen0
    {%
      \hss
      \vrule width \epsfframethickness
      \kern \epsfframemargin
      \vbox {\kern \epsfframemargin \box0 \kern \epsfframemargin }%
      \kern \epsfframemargin
      \vrule width \epsfframethickness
      \hss
    }
    \hrule height 0pt depth \epsfframethickness
  }
}%
\def\epsfsetgraph#1%
{%
   %
   %
   \leavevmode
   \hbox{
     \ifepsfframe\expandafter\epsfframe\fi
     {\vbox to\epsfysize
     {%
        \ifepsfshow
            \vfil
            \hbox to \epsfxsize{\epsfspecial{#1}\hfil}%
        \else
            \vfil
            \hbox to\epsfxsize{%
               \hss
               \ifepsfshowfilename
               {%
                  \epsfframemargin=3pt 
                  \epsfframe{{\tt #1}}%
               }%
               \fi
               \hss
            }%
            \vfil
        \fi
     }%
   }}%
   %
   %
   \global\epsfxsize=0pt
   \global\epsfysize=0pt
}%
%
%
\def\epsfsetsize
{%
   \epsfrsize=\epsfury\pspoints
   \advance\epsfrsize by-\epsflly\pspoints
   \epsftsize=\epsfurx\pspoints
   \advance\epsftsize by-\epsfllx\pspoints
%
%
   \epsfxsize=\epsfsize{\epsftsize}{\epsfrsize}%
   \ifnum \epsfxsize=0
      \ifnum \epsfysize=0
        \epsfxsize=\epsftsize
        \epsfysize=\epsfrsize
        \epsfrsize=0pt
%
%
      \else
        \epsftmp=\epsftsize \divide\epsftmp\epsfrsize
        \epsfxsize=\epsfysize \multiply\epsfxsize\epsftmp
        \multiply\epsftmp\epsfrsize \advance\epsftsize-\epsftmp
        \epsftmp=\epsfysize
        \loop \advance\epsftsize\epsftsize \divide\epsftmp 2
        \ifnum \epsftmp>0
           \ifnum \epsftsize<\epsfrsize
           \else
              \advance\epsftsize-\epsfrsize \advance\epsfxsize\epsftmp
           \fi
        \repeat
        \epsfrsize=0pt
      \fi
   \else
     \ifnum \epsfysize=0
       \epsftmp=\epsfrsize \divide\epsftmp\epsftsize
       \epsfysize=\epsfxsize \multiply\epsfysize\epsftmp
       \multiply\epsftmp\epsftsize \advance\epsfrsize-\epsftmp
       \epsftmp=\epsfxsize
       \loop \advance\epsfrsize\epsfrsize \divide\epsftmp 2
       \ifnum \epsftmp>0
          \ifnum \epsfrsize<\epsftsize
          \else
             \advance\epsfrsize-\epsftsize \advance\epsfysize\epsftmp
          \fi
       \repeat
       \epsfrsize=0pt
     \else
       \epsfrsize=\epsfysize
     \fi
   \fi
}%
%
%
\def\epsfstatus#1{
   \ifepsfverbose
     \immediate\write16{#1: BoundingBox:
                  llx = \epsfllx\space lly = \epsflly\space
                  urx = \epsfurx\space ury = \epsfury\space}%
     \immediate\write16{#1: scaled width = \the\epsfxsize\space
                  scaled height = \the\epsfysize}%
   \fi
}%
%
%
{\catcode`\%=12 \global\let\epsfpercent=
\global\def\epsfatend{(atend)}%
%
%
%
%
%
%
%
\long\def\epsfaux#1#2:#3\\%
{%
   \def\testit{#2}
   \ifx#1\epsfpercent           
       \ifx\testit\epsfbblit    
            \epsfgrab #3 . . . \\%
            \ifx\epsfllx\epsfatend 
                \global\epsfatendtrue
            \else               
                \ifepsfatend    
                \else           
                    \epsffileokfalse
                \fi
                \global\epsfbbfoundtrue
            \fi
       \fi
   \fi
}%
%
%
\def\epsfempty{}%
\def\epsfgrab #1 #2 #3 #4 #5\\{%
   \global\def\epsfllx{#1}\ifx\epsfllx\epsfempty
      \epsfgrab #2 #3 #4 #5 .\\\else
   \global\def\epsflly{#2}%
   \global\def\epsfurx{#3}\global\def\epsfury{#4}\fi
}%
%
%
\def\epsfsize#1#2{\epsfxsize}%
%
%
\let\epsffile=\epsfbox